\documentclass[%
 reprint,
superscriptaddress,
nofootinbib,
 amsmath,amssymb,
 aps,
]{revtex4-2}
\usepackage{amsthm}
\usepackage{color}
\usepackage{xcolor}
\usepackage{mathrsfs}
\newtheorem*{remark}{Conjecture}
\usepackage{graphicx}
\usepackage{dcolumn}

\usepackage{bm}

\begin{document}

\preprint{APS/123-QED}

\title{Probing of Colliding Wave Spacetimes and a Quantum Weyl Conjecture}

\author{Ivo Sachs}
\email{ivo.sachs@lmu.de}
 \affiliation{Arnold-Sommerfeld Center for Theoretical Physics,\\
Theresienstra{\ss}e 37, 80333 München, Germany}
\author{Marc Schneider}%
 \email{marc.schneider@ehu.eus}
\affiliation{%
 Department of Physics and EHU Quantum Center, University of the Basque Country UPV/EHU,\\
Barrio Sarriena s/n, Leioa 48940, Spain
}%

\date{\today}

\begin{abstract}
We perform a quantum probing of colliding plane-wave space-times. In particular, we consider the Khan-Penrose and the Ferrari-Ibáñez solutions, which admit a strong and a weak singularity after the two waves collide. While we find that, like Schwarzschild, for the Khan-Penrose solution the singularity cannot be probed by quantum field theory, the Ferrari-Ibáñez singularity can be traversed. Our results culminate in a \emph{quantum Weyl conjecture}: The significant geometric property to classify space-times with respect to quantum probes is given by the Coulomb part of the Weyl tensor. We then use this conjecture to sketch a possible backreaction scenario for plane waves.
\end{abstract}

\maketitle


\section{Introduction}
\label{sec:intro}

Quantum probing revealed that the intuitive classical picture fails when quantized probes are considered. The concept contrasts the occurrence of a singularity with its physical significance. Colloquially speaking, addressing the question: Do singular systems lead to singular measurements? The answer is in the negative and one of the showcase examples marks the hydrogen atom. Albeit incomplete with respect to classical electrodynamics, quantum mechanical evolution remains complete because of a probabilistic completion, i.e. there is a vanishing probability to find  bound-states, as well as scattering states of the electron to be found at the center, where the Coulomb field becomes singular. Hence, the electron is barred from probing the endpoint. 

The concept of quantum mechanical completeness has been extended to quantum field theory in curved space-times to capture dynamical scenarios in which absorption and emission processes occur \cite{hof15}. One of the most intricate questions is the fate of quantum fields that encounter curvature singularities. The functional Schrödinger representation has proven itself to be extremely versatile to detect pathologies in a time-dependent framework. For the Schwarzschild black hole's interior, one observed a probabilistic suppression that led to a vanishing probability of finding any quanta on the singular hypersurface \cite{egl17}. In this sense, Schwarzschild black holes comply with the idea of being gravitational analogs of the hydrogen atom; however, the discussion here focuses on scattering states while bound states are not considered.
In fact, our functional Schrödinger states describe the evolution of such field configurations and thus determine the probability density of a quantum field to be located on a particular hypersurface through the norm.

While the gravitational singularity that occurs inside Schwarzschild black holes seems to prevent the fields from reaching it \cite{hof15}, the focusing singularity in plane-wave space-times shows a different behavior. In fact, after passing through the wave, the quantum fields experience a focusing and will eventually reach the focal plane \cite{sachs21}. The different quality attributed to the singularity influences the behavior of the quantum fields directly. The obvious question arises: what exactly is this quality?

To answer this question, we consider colliding plane wave space-times because they offer a rich geometric playground that combines plane waves and black hole interiors \cite{khan,ferrari1987new,szekeres1970colliding,chandrasekhar1985colliding,yurtsever1988new,hayward1989colliding}. This class of space-times describes colliding plane waves which form a very diverse geometric landscape with four patches that are separated by the sandwich waves. Before the collision, one faces a  flat region (denoted below as patch I), which resembles Minkowski space-time except for the future boundaries which are the plane waves. After crossing one of the incoming waves, there lies a region with a focusing, null singularity (patches II and III). These regions correspond to pp-wave space-times \cite{pen65}. The most intriguing region is patch IV, which forms after the waves have collided. There, the collision of the waves creates a scatter that propagates towards the future but is destined to collapse under its own gravitational pull, thus forming a spacelike singularity. There exist several solutions depending on the details of the system from which we choose to work with the Khan-Penrose solution \cite{khan}. In this particular solution, the curved region features a gravitational collapse reminiscent to the formation of a black hole and, hence, amounts to a Schwarzschild interior \cite{yurtsever1988structure,yurtsever1988colliding}. However, there are milder solutions that admit weaker singularities after the collision, e.g. the Ferrari-Ibáñez space-time \cite{ferrari1987new}, which we study for reference as well.

Since we are interested in the general properties that determine the evolution of quanta on curved backgrounds, (colliding) plane-wave geometries are ideal because every space-time provides locally a plane-wave limit \cite{blau2004penrose,feinstein2002penrose}. Therefore, probing all patches of this geometry allows us to link the folding singularities of the focal plane with the curvature singularity after the collision and extract the geometric properties that determine the fate of quantum probes. Regarding the classification of the geometric part, we will make use of the Penrose-Newman formalism \cite{penrose1986spinors} which allows to decompose the different parts of the curvature tensor. These Penrose-Newman scalars provide a clear interpretation and allow us to decode the properties of gravity that determine the evolution of the quantum probes. In the underlying case, a strong connection with the Coulomb part of the Weyl tensor emerges, as such, substantiating the intuition of black holes being gravitational analogs of the hydrogen atom. We will formalize this observation into a \emph{quantum Weyl conjecture}. 

Our analysis will be organized as follows: In the next section, we discuss the geometric properties of colliding plane wave space-times, that is, the different regions shaped by the waves; our main focus will be the region after the collision. Then, in the third section, we discuss the geometric properties of the Khan-Penrose solution explicitly and perform a quantum probing of patch IV. This we contrast with the Ferrari-Ibáñez solution in section 4. We collect our findings and weave them into the quantum Weyl conjecture in section 5 where we furthermore apply this conjecture to estimate a (quasi-classical) backreaction scenario. We close with a discussion of our results in the last section. Throughout the article, we use the mostly plus sign convention for the metric and natural units.

\section{The geometry of colliding plane waves}
\label{sec:2}

In this section, we describe the basic geometric facts about plane-wave space-times and their properties after collision. Plane-wave space-times have a wide range of applications and constitute even a particular limit to many space-times that seem initially not connected \cite{pen76}. We start with the description of single plane-waves and move towards the case where they collide. 

\subsection{Plane-wave space-time}

To describe a collision of two plane waves, we begin with a single, linearly polarized plane wave with profile $H(u,x,y)$ that travels in the $v$-direction \cite{pen65,garr91}. In Brinkmann coordinates,
\begin{equation}\label{eq:gBrinkmann}
    g=-2\mbox{d}u\otimes\mbox{d}v-H(u,x,y)\mbox{d}u\otimes\mbox{d}u+\mbox{d}x\otimes\mbox{d}x+\mbox{d}y\otimes\mbox{d}y\,,
\end{equation}
where $u$ and $v$ denote the two null directions. This patch describes how the wave propagates on an otherwise flat Minkowski space-time. If the plane wave is sourced by matter, the latter enters through the stress-energy tensor
\begin{equation}
    T_{uu}=-\frac{1}{8\pi}\mbox{tr}_g(H(u,x,y))
\end{equation}
 into the Einstein equations. Albeit intuitive from an overseeing perspective, the more instructive description of the dynamics provides the Einstein-Rosen patch. Through the coordinate transformations $u\to U$, $v\to V+\frac{X^2}{2} F(U)\frac{{\rm d}F}{{\rm d}U}(U)+\frac{Y^2}{2} G(U)\frac{{\rm d}G}{{\rm d}U}(U)$, $x\to F(U)X$, and $y\to G(U)Y$, the metric \eqref{eq:gBrinkmann} becomes
\begin{equation}\label{eq:gERosen}
g_{III}=-2\mbox{d}U\otimes\mbox{d}V+F^2(U)\mbox{d}X\otimes\mbox{d}X+G^2(U)\mbox{d}Y\otimes\mbox{d}Y.
\end{equation}
The two functions $F(U)$ and $G(U)$ depend on the specifics of the wave profile $H(U,x,y)$ through the equations 
\begin{equation}\label{eq:planewavesingle}
\frac{{\rm d}^2F}{{\rm d}U^2}=H(U)F(U),\quad\mbox{and}\quad \frac{{\rm d}^2G}{{\rm d}U^2}=-H(U)G(U),
\end{equation}
which for an impulsive wave of unit amplitude, that is $H(U)\propto\delta(U)$, become $F(U)=1+U\,\Theta(U)$ and $G(U)=1-U\,\Theta(U)$ where $\Theta(U)$ denotes the Heaviside distribution. 

From, \eqref{eq:planewavesingle} we then infer, in particular, that  null-geodesics in $U$-direction, originating at past null-infinity, experience a flat space-time unless they have crossed the wave where they are diverted and focused on a null-singularity at $U=1$. The same type of space-time can be constructed when the wave travels in the $u$-direction instead. Then, the metric functions become $V$-dependent
\begin{equation}\label{eq:focusV}
g_{II}=-2\mbox{d}U\otimes\mbox{d}V+F^2(V)\mbox{d}X\otimes\mbox{d}X+G^2(V)\mbox{d}Y\otimes\mbox{d}Y,
\end{equation}
where the functions $F(V)$ and $G(V)$ can be distinct from the ones in \eqref{eq:planewavesingle}, because they depend on the profile and strength of the wave in $U$-direction. This can analogously be transformed into the Brinkmann coordinates via choosing an appropriate profile $H(v,x,y)$ for the wave that propagates in the $u$-direction. 

\subsection{Colliding plane waves}

A collision of two impulsive waves combines the two previously discussed scenario and, therefore,  admits the space-time $(\mathbb{K},g)$, i.e. a manifold $\mathbb{K}$ and a metric $g$ which in Einstein-Rosen coordinates reads (with parallel linear polarization) \cite{yurtsever1988structure}
\begin{equation}\label{eq:yurtsevermetrikform}
    g=-2e^{-M(U,V)}\mbox{d}U\otimes\mbox{d}V+\gamma.
\end{equation}
The spatial part is given through the $2\times2$-metric
\begin{equation}
    \gamma=e^{-\beta(U,V)}(e^{\alpha(U,V)}\mbox{d}X\otimes\mbox{d}X+e^{-\alpha(U,V)}\mbox{d}Y\otimes\mbox{d}Y)
\end{equation}
where $\alpha$ and $\beta$ are functions of $U$ and $V$ only. This space-time admits a planar symmetry in the $X$-$Y$ plane. There exists a canonical null tetrad which consists of the null vectors $l$ and $n$ that are tangent to the null geodesic congruences and perpendicular to the two Killing vectors $\xi_X$ and $\xi_Y$ \cite{penrose1986spinors,yurtsever1988structure}. These form the remaining basis vectors $m$ and its complex conjugate $\bar m$. For parallel linear polarizations, they admit the fairly simple form
\begin{eqnarray}\label{eq:frame}
    l&=&2\;e^M\frac{\partial}{\partial U},\quad n=\frac{\partial}{\partial V}\nonumber,\\
    m&=&\frac{1+i}{2}\;e^\frac{\beta-\alpha}{2}\frac{\partial}{\partial X}+\frac{1-i}{2}\;e^\frac{\beta+\alpha}{2}\frac{\partial}{\partial Y},\\
     \bar m&=&\frac{1-i}{2}\;e^\frac{\beta-\alpha}{2}\frac{\partial}{\partial X}+\frac{1+i}{2}\;e^\frac{\beta+\alpha}{2}\frac{\partial}{\partial Y}\nonumber.
\end{eqnarray}
These vectors play a central role in the construction of the curvature scalars that we exploit to understand the properties of the possible singularity. 

Considering now the collision of two parallel linear polarized waves. Depending on the functions $M$, $\alpha$, and $\beta$, the metric describes following four portions: 
\begin{enumerate}
    \item[(I)] a flat region, before any wave has been crossed, that is, $M=\alpha=\beta\equiv 0$, thus, the metric becomes the Minkowski metric
    \item[(II)] the focusing regions behind the wave in $U$-direction that focuses in $V=1$ for which $M=0$, $\alpha=\alpha(V)$, and $\beta=\beta(V)$ and the metric reads \eqref{eq:focusV}
    \item[(III)] the focusing regions behind a wave in $V$-direction that focuses in $U=1$ for which $M=0$, $\alpha=\alpha(U)$, and $\beta=\beta(U)$ and the metric reads \eqref{eq:planewavesingle}
    \item[(IV)] the curved region that forms after the plane waves have collided for which $M=M(U,V)$, as well as $\alpha=\alpha(U,V)$ and $\beta=\beta(U,V)$. 
\end{enumerate}
To understand the asymptotic structures of all these patches, the Penrose-Newman scalars serve as a versatile tool. Since we are dealing with vacuum solutions, the relevant curvature stems from the five Weyl scalars, defined via a contraction of the Weyl curvature tensor $C$ with the canonical frame fields. Using the metric \eqref{eq:yurtsevermetrikform} and the frame fields \eqref{eq:frame} we find \cite{penrose1986spinors}
\begin{eqnarray}\label{eq:WeylPN}
    \Psi_0&=&-C(l,m,l,m)\nonumber\\
    &&=2 i\;e^{2M}(\partial_UM\;\partial_U\alpha+\partial_U^2\alpha-\partial_U\alpha\;\partial_U\beta) ,\nonumber\\
    \Psi_1&=&-C(l,m,l,n)=0,\nonumber\\
    \Psi_2&=&-C(l,m,\bar m,n)=-e^M\;\partial_U\partial_VM,\\
    \Psi_3&=&-C(l,n,\bar m,n)=0,\nonumber\\
    \Psi_4&=&-C(\bar m,n,\bar m,n)\nonumber\\
    &&=\frac i2(\partial_V\alpha\;\partial_V\beta-\partial_VM\;\partial_V\alpha-\partial_V^2\alpha)\nonumber.
\end{eqnarray}
The $\Psi_0$ and $\Psi_4$ scalars describe contributions of transverse waves propagating in the $n$- or $l$-direction respectively while $\Psi_1$ and $\Psi_3$ denote the respective longitudinal contributions; the scalar $\Psi_2$ is particular as a Coulomb like contribution because it contains both, derivatives with repect to $U$ and $V$ while the other scalar only consist of one sort of derivative. 

The Weyl scalars allow us to relate the patches of the plane wave space-time with the Petrov classification of space-times, so are the focusing regions of type N since all but either $\Psi_0$ or $\Psi_4$ vanish while the region before any wave has been crossed is of type O. The curved patch itself does not fit into the Petrov classification but asymptotically, some solutions resemble a Petrov type D space-time because all except $\Psi_2$ vanish (or grow slower than $\Psi_2$). Since plane-waves generically lead to a focusing of geodesics, these space-times will be geodesically incomplete, i.e. admit singularities \cite{hawk73}. We will see, how this affects our results eventually.

In the following, we focus on two particular solutions with different strength of singularities: the Khan-Penrose solution and the Ferrari-Ibáñez solution  \cite{yurtsever1989singularities,yurtsever1988structure}.

\section{Khan-Penrose space-time}

In this section, we consider the collision of two impulsive null-waves ($\delta$-distributional profile) that is described by the Khan-Penrose space-time \cite{khan}. We describe, both its geometrical properties and the implication for the quantum probes.

\subsection{Khan-Penrose space-time: geometry}
The particular form of such a space-time has been studied by Khan and Penrose in \cite{khan} and found to be of the form
\begin{align}\label{eq:khanpenrose}
    g=-2&e^{-M(U,V)}\mbox{d}U\otimes\mbox{d}V\\
&\hspace{1em}+F^2(U,V)\mbox{d}X\otimes\mbox{d}X+G^2(U,V)\mbox{d}Y\otimes\mbox{d}Y\nonumber
\end{align}
where the metric coefficients are determined by the three functions \begin{eqnarray}
   e^{-M(U,V)}&=&\frac{s(U,V)^\frac{3}{2}}{w(U)w(V)(p(U)q(V)+w(U)w(V))^2},\;\;\\
    F^2(U,V)&=&s(U,V)\frac{1-p(U)w(V)-q(V)w(U)}{1+p(U)w(V)+q(V)w(U)},\\
    G^2(U,V)&=&s(U,V)\frac{1+p(U)w(V)+q(V)w(U)}{1-p(U)w(V)-q(V)w(U)}.
\end{eqnarray}
For the ease of notation, we have defined the three quantities $w(V)=\sqrt{1-q(V)^2}$,  $w(U)=\sqrt{1-p(U)^2}$, and $s(U,V)=1-p(U)^2-q(V)^2$. 

\begin{figure}
    \centering
\includegraphics[width=\linewidth]{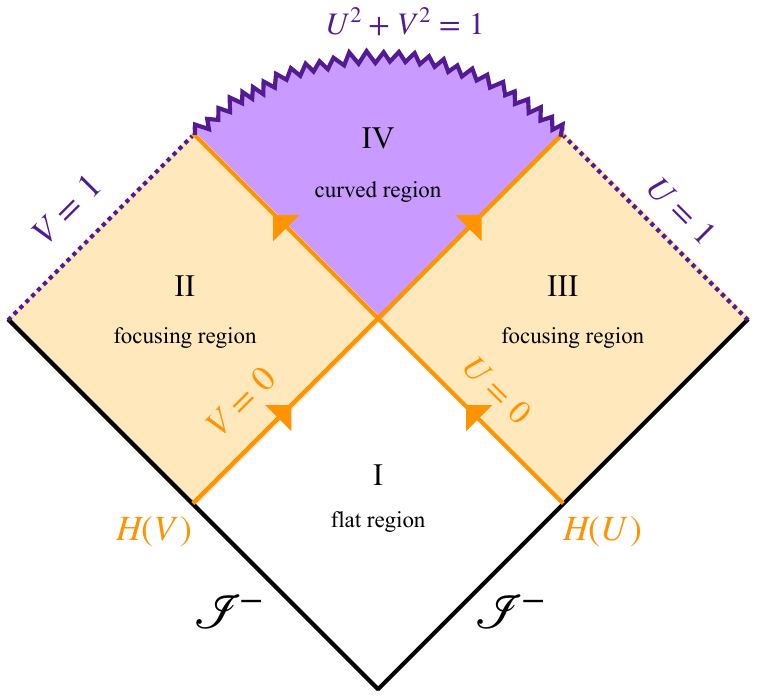}
    \caption{Penrose diagram of the Khan-Penrose space-time showing two colliding waves at $U=0$ and $V=0$ that cut the space-time into four portions. In the orange shaded, focusing regions, null-geodesics that have crossed the wave are focused at the focal plane $V=1$ in II, or $U=1$ in III. These mark the folding singularities. The purple shaded region is curved and admits a space-like curvature singularity at $U^2+V^2=1$. 
    \label{fig:KPRZl2}}
\end{figure}

By adopting the impulsive wave profile, we can derive the explicit forms of the functions $p(U)=U\,\Theta(U)$ and $q(V)=V\,\Theta(V)$. As such, the two waves propagate along $U=0$ and the other along $V=0$, thus cutting the manifold into four portions (cf. fig. \ref{fig:KPRZl2}):
\begin{itemize}
    \item before crossing any wave: for $U<0$ and $V<0$, the three space-time functions $e^{-M(U,V)}=F^2(U,V)=G^2(U,V)\equiv1$ and \eqref{eq:khanpenrose} reduces to Minkowski space-time in light-cone coordinates.
    \item crossing one wave: for either $U<0$ and $V>0$ or $U>0$ and $V<0$ the space-time is flat again but geodesics will be focused at the null-singularity at either $V=1$ or $U=1$ respectively (or $w=0$). 
    \item region after the collision: for $U>0$ and $V>0$, the space-time becomes curved and any geodesic will end in a space-like curvature singularity at $U^2+V^2=1$ (or $s(U,V)=0$). 
\end{itemize}
As we can see, the singularity cannot be avoided by any geodesic and, therefore, the space-time is g-incomplete \cite{hawk73} and not even globally hyperbolic due to the focusing singularities that completely enclose the future development \cite{pen65}.

Since we are mostly interested in the curved region, we will focus on patch IV\footnote{It is clear that by replacing $U$ with the function $p(U)$ and $V$ with $q(V)$, we can chart the whole space-time.} except in section \ref{sec:5}  we will consider backreaction. The curvature properties and in particular the singular asymptotics can be understood by the non-zero components of the Weyl tensor
\begin{eqnarray}
    &&\Psi_0=\frac{\delta(V)}{w(U)}-\frac{3U\,w(U)(w(U)w(V)+UV)}{w(V)^2s(U,V)^2},\label{eq:psi0}\\
    &&\Psi_2=\frac{(UV+w(U)w(V))^2}{w(U)w(V)s(U,V)^2}-\frac{UV}{s(U,V)^2},\label{eq:psi2}\\
    &&\Psi_4=\frac{\delta(U)}{w(V)}-\frac{3V\,w(V)(w(U)w(V)+UV)}{w(U)^2s(U,V)^2},\label{eq:psi4}
\end{eqnarray}
which obviously diverge whenever $U^2+V^2=1$. For the ease of notation, we have defined the quantities $w$ and $s$ as above. Note that the Kretschmann scalar can be composed out of the Weyl scalars via
\begin{equation}\label{eq:Kretschmann}
    K=R^{abcd}R_{abcd}= \mbox{Re}(3\Psi_2^2+\Psi_0\Psi_4)
\end{equation}
where $R_{abcd}$ denotes the Riemann curvature tensor.
Since the Kretschmann scalar is a curvature invariant, the corresponding singularity is not removable. Similar to Schwarzschild space-time, $\Psi_2$ diverges at least as fast as the squared reciprocal; such a singularity is considered to be strong in the sense of Tipler and Kr\'olak \cite{tipler1978energy,clarke1985conditions}. From a physics perspective, this region can be understood as the scattered field that is produced by the collision which eventually suffers self-focusing. In this region, also $\Psi_0$ and $\Psi_4$ diverge in the same manner. 

It should be mentioned that the folding singularity in the patches II and III does not fulfill this criterion because, in both, $\Psi_2\equiv0$ and either $\Psi_4=0$ (in II) or $\Psi_0=0$ (in III) while the corresponding other scalar remains non-zero, even diverges. Their main divergence can be read off from \eqref{eq:psi0} and \eqref{eq:psi4} to scale with the reciprocal only, that is $(1-V^2)^{-1}$ in II and $(1-U^2)^{-1}$ in III. Since the condition for a strong singularity has not been met the folding singularity is considered to be weak in the sense of Tipler and Kr\'olak \cite{yurtsever1988structure}.

Considered separately, region IV is a globally hyperbolic space-time for a spatial slicing that mimics the Schwarzschild interior. For convenience, we define a temporal and a spatial coordinate perpendicular to the symmetry plane as follows \cite{garr91}
\begin{equation}
\xi=U+V\quad\mbox{and}\quad\eta=U-V.\label{eq:1+3splitcoords}    
\end{equation}
while $X\to X$ and $Y\to Y$. Here, $\xi$ defines a time while $\eta$ denotes a spatial coordinate in such a way that the region IV is foliated by Cauchy surfaces and the past asymptotic boundaries are determined by the impulsive waves. Additionally, in the coordinate chart $(\xi,\eta,X,Y)$ the metric assumes a diagonal form that facilitates the functional Schrödinger analysis. 

\subsection{Khan-Penrose space-time: quantum probing}

To study the fate of quantum probes, we work in the functional Schrödinger representation of quantum field theory which treats the evolution of entire field configurations rather than the individual fields. The functional Schrödinger approach is based on the canonical phase space description which presumes a well-defined Cauchy evolution (or a null-evolution) to be complete \cite{hofmann25}. The Khan-Penrose space-time naturally offers the possibility to construct the phase space either in dual-null or the temporal-spatial foliation \cite{yurtsever1989quantum} depending on the geometry of the individual region. For instance while the singularity structure in region II and III favors a dual null foliation, region IV clearly suggests a timelike foliation of spacelike hypersurfaces. 

The quantum probing of the null singularity in II and III has been performed previously in \cite{sachs21} with the result that the folding singularity is reachable and the quantum field passes the focal plane although in section \ref{sec:5} we will argue that back reaction becomes important in that region. However, in this section, our main focus rests upon the singularity in the curved region, since this region serves as a Penrose limit to a Schwarzschild black hole. 

We start with a review of the temporal-spatial split that is adapted to the geometry of the collision region. We construct the Hamilton operator through a temporal-spatial split \cite{hayw93}: consider the configuration space, parametrized by the field $\phi$ pulled back to the spatial hypersurface $\Sigma$ and $(\phi,\pi)\in T^*\mathcal{F}\Sigma$, where  $\mathcal{F}\Sigma$ denotes the space of smooth functions on the spatial hypersurface $\Sigma$.
The Lagrange density $\mathcal{L}$ defines a Lagrange transformation $\Lambda:T\mathcal{F}\Sigma\to T^*\mathcal{F}\Sigma$, with $\Lambda:(\phi,\dot \phi)\mapsto(\phi,\frac{\delta\mathcal{L}}{\delta\dot{\phi}})=(\phi,\pi)$. 
Now, we construct the $(1+3)$-foliation by using the coordinate chart $(\xi,\eta,X,V)$, where $\xi$ and $\eta$ are defined in \eqref{eq:1+3splitcoords}. Then, for a massless, minimally coupled, scalar field, the Hamilton operator becomes
\begin{equation}\label{eq:hamiltonOp}
    \mathcal{H}=\frac{\sqrt{-g_{\xi\xi}}}{2}\left(\frac{\pi^2}{{\rm det}(\sigma)}+\sigma(\nabla\phi,\nabla\phi)\right)
\end{equation}
where we defined the metric $\sigma$ as the induced metric on the spatial hypersurface $\Sigma$. Note, the operators $\phi$ are the configuration variable in the field configuration space while the momentum conjugate to $\phi$ is given by
\begin{equation}
    \pi=\frac{\delta\mathcal{L}}{\delta\dot\phi}=\frac{\sqrt{{\rm det}(\sigma)}}{\sqrt{-g_{\xi\xi}}}\dot\phi\,,
\end{equation}
where the overdot-notation symbolizes the derivative with respect to $\xi$. The Hamilton operator rules the dynamics of the field configurations through a functional Schrödinger equation
\begin{equation}\label{eq:SGL}
i\partial_\xi\Psi_\xi[\phi]=H_\xi[\pi,\phi]\Psi_\xi[\phi],
\end{equation}
and the wave-functional $\Psi_\xi:\mathcal{F}\Sigma_\xi\to\mathbb{C}$, hence, describes the evolution of the field configurations that are pulled back on $\Sigma_\xi$. By imposing the canonical quantization prescription $[\pi(x),\phi(x')]=i\delta(x,x')$ we can express $\pi\to-i\frac{\delta}{\delta\phi}$ as the functional derivative operator while $\phi$ becomes the multiplicative operator $\phi|\Psi\rangle=\langle\phi\rangle_{\rm cl}|\Psi\rangle$ which yields the classical expectation value $\langle\phi\rangle_{\rm cl}=:\phi$ as eigenvalue.

The functional Schrödinger states live in the  $\mathscr{L}^2(\mathcal{F}\Sigma_\xi,\mathscr{D}\phi)$-Hilbert space. Note, $\mathscr{D}\phi$ denotes a Lebesgue like measure over all field configurations. In this formal measure space, the inner product is defined by integrating over all field configurations $\phi\in \mathcal{F}\Sigma_\xi$
\begin{equation}\label{eq:FSSNorm}
\|\Psi_\xi\|^2=\int_{\mathcal{F}\Sigma_\xi}\mathscr{D}\phi\;\left|\Psi_\xi[\phi]\right|^2.
\end{equation}
This functional integral can be interpreted as the probability to populate a certain hypersurface $\Sigma_\xi$ \cite{hof19}. This is the quantity we use to measure the probability for field quanta to be present on a spatial hyper surface of space-time. In particular we are interested how this probability evolves as the hypersurface approaches the singularity. If this probability vanishes the corresponding region is not accessible to a quantum field. On the other hand if it diverges back reaction should be taken into account.

The geometries we consider below approach Minkowski space as $\xi\to -\infty$. Accordingly the $in$ vacuum can be defined unambiguously \cite{yurtsever1989quantum}. As an initial condition we then consider a plane wave over this state. The fact that we consider a free field theory, power counting arguments using \eqref{eq:hamiltonOp} suggest a Gaussian state for the wave-functional
\begin{equation}\label{eq:SchrödingerWF}
\Psi_\xi[\phi]=N(\xi)\;\mathcal{G}_\xi[\phi]\hspace{0.7em}\mbox{with}\hspace{0.7em} \mathcal{G}_\xi[\phi]=\exp\left(-\frac12[\phi]\mathcal{K}(\xi)[\phi]\right).
\end{equation}
We observe the two distinct contributions, a field-configuration dependent exponential $\mathcal{G}_\xi[\phi]$ and a field independent normalization $N(\xi)$. The functional kernel $\mathcal{K}_\xi:\mathcal{F}\Sigma_\xi\otimes \mathcal{F}\Sigma_\xi\to\mathbb{C}$
\begin{equation}\label{eq:kernelallg}
[\phi]\mathcal{K}_\xi[\phi]=\iint_{\Sigma_\xi}{\rm d}^3\mu(x,x')\phi(x)K(x,x';\xi)\phi(x')
\end{equation}
inherits a time-dependence from the metric. The measure d$^3\mu(x,x')=\;$d$^3x$d$^3x'\sqrt{{\rm det}(\sigma)(x)}\sqrt{{\rm det}(\sigma)(x')}$. We will now see that the bi-local kernel function $K(x,x';\xi)$ completely determines the dynamics of the field configurations.

With the Gaussian ansatz \eqref{eq:SchrödingerWF}, the norm factorizes $\|\Psi_\xi\|^2=|N(\xi)|^2\;\|\mathcal{G}_\xi[\phi]\|^2$ into the modulus square of the time-dependent function $N(\xi)$ and the field dependent functional integral over $\mathcal{G}_\xi[\phi]$. Through insertion of \eqref{eq:SchrödingerWF} into the functional Schrödinger equation \eqref{eq:SGL}, one finds the normalization to be
\begin{equation}
 N(\xi)=N_0\;\exp\left(-\frac i2\int_\mathbb{K}\sqrt{-{\rm det}(g)}K(x,x;\xi)\right),
\end{equation}
where we defined the constant of integration $N_0$ which encodes the initial conditions. The planar symmetry in $\mathbb{K}$ allows us to perform a Fourier transformation in the spatial coordinates. Thus, the first contribution to the norm, $|N(\xi)|^2$, is evaluated as follows 
\begin{equation}\label{eq:FSSNNorm}
    \frac{|N(\xi)|^2}{|N_0|^2}=\exp\left(\int_\mathbb{K}\mbox{d}^4x\int_k\frac{{\rm d}\boldsymbol k}{(2\pi)^2}\sqrt{-\mbox{det}(g)}\,\mbox{Im}(K_{\boldsymbol{k}})(\xi)\right).
\end{equation}
We observe, that the normalization depends on the imaginary part of the kernel while the norm of the field dependent part contains its real part
\begin{equation}
\|\mathcal{G}_\xi[\phi]\|^2=\int\mathscr{D}\phi\;\exp\left(-[\phi]{\rm Re}(\mathcal{K}_\xi)[\phi]\right).
\end{equation}
When plugging \eqref{eq:SchrödingerWF} into \eqref{eq:SGL}, one obtains a Riccati equation for $K(x,x';\xi)$ which, in turn, can be transformed into the Klein-Gordon equation. As such, we can construct $K(x,x';\xi)$ from the mode solutions \cite{long1998schrodinger,hof15}.

Without loss of generality \cite{hof17}, we may assume our probe theory to be a free, minimally coupled, massless scalar field $\phi(x)$ that obeys the Klein-Gordon equation $\Box\phi(x)=0$. As we see in figure \ref{fig:KPRZl}, there are essentially three distinct regions where the field propagates quite differently. Since the asymptotic regions are flat space-time, there exists a natural vacuum such that the rightgoing modes (along $U$) are Minkowski solutions in I and II, while the leftgoing modes (along $V$) are Minkowski in I and III \cite{yurtsever1989quantum}. 

However, our main interest lies on the behavior in the curved patch, and we will, therefore, construct the functional Schrödinger representation of our quantum field theory only there. Next, let us solve the Klein-Gordon equation in region IV using the metric \eqref{eq:khanpenrose}. For the general form of the metric, the Klein-Gordon equation in dual null-coordinates $(U,V,X,Y)$ reads \cite{yurtsever1989quantum}
\begin{align}\label{eq:KGd-0-coords}
    \Box\phi=-e^{M(U,V)}&\bigg[2\partial_U\partial_V\phi+\partial_U\ln(FG)\partial_V\phi\\
&+\partial_V\ln(FG)\partial_U\phi\bigg]+\frac{\partial_X^2\phi}{F^2}+\frac{\partial_Y^2\phi}{G^2}=0\nonumber.
\end{align}
We are interested in particular in the behavior close to the singularity, that is, for $U^2+V^2\to1$. 
The form of the metric components is rather complicated and since our objective is to probe the singularity, we approximate the metric close to the singularity and find
\begin{align}\label{eq:asympt-dual-0}
    g\simeq&-\frac{(1-(U^2+V^2))^{\frac32}}{U^2V^2}\mbox{d}U\otimes\mbox{d}V\\
    &+\frac{(1-(U^2+V^2))^2}{2}\mbox{d}X\otimes\mbox{d}X+2\mbox{d}Y\otimes\mbox{d}Y.\nonumber
\end{align}
It is obvious that the null directions as well as the $X$-direction vanish as we approach the singularity,  while the $Y$-direction assumes a finite sized value. The planar symmetry is reflected in the Killing vector fields $\partial_X$ and $\partial_Y$, which allows us to perform a Fourier decomposition 
\begin{equation}\label{eq:Fourierzerlegung}
    \phi(U,V,X,Y)=\iint\frac{{\rm d}^2\boldsymbol{k}}{4\pi^2}\chi_{\boldsymbol{k}}(U,V)e^{i\boldsymbol{k}\boldsymbol{x}} ,
\end{equation}
where $\boldsymbol{k}=(k_X,k_Y)$ and $\boldsymbol{x}=(X,Y)^T$.
With the asymptotic metric \eqref{eq:KGd-0-coords} and the Fourier decomposition \eqref{eq:Fourierzerlegung}, the Klein-Gordon equation becomes, in the near singularity limit, 
\begin{align}\label{eq:KG-asy-dual0}
    &\frac{2U^2V^2}{\sqrt{1-(U^2+V^2)}}\partial_U\partial_V\chi+U\partial_V\chi+V\partial_U\chi\\
    &+\left(\frac{k^2_X}{1-(U^2+V^2)}+(1-(U^2+V^2))k^2_Y\right)\chi=0\,\nonumber.
\end{align}
Note that we suppressed the dependencies of $\chi$ for ease of notation.

Unfortunately, the coefficients in the differential equation prohibit a factorization due to the, still, complicated mixing of $U$ and $V$. Therefore, we change to the coordinate chart \eqref{eq:1+3splitcoords} to naturally define the phase space. In this coordinates, the metric reads
\begin{align}\label{eq:metrik-xi-eta}
   g\simeq- &\frac{|1-\frac{\xi^2+\eta^2}{2}|^{\frac32}}{(\xi^2-\eta^2)^2}\mbox{d}\xi\otimes\mbox{d}\xi+ \frac{|1-\frac{\xi^2+\eta^2}{2}|^{\frac32}}{(\xi^2-\eta^2)^2}\mbox{d}\eta\otimes\mbox{d}\eta\\
   &+\frac{|1-\frac{\xi^2+\eta^2}{2}|^2}{2}\mbox{d}X\otimes\mbox{d}X+2\mbox{d}Y\otimes\mbox{d}Y\nonumber
\end{align}
From here, we can construct the d'Alembert operator accordingly in its explicit (asymptotical) form. Using again the planar symmetry to Fourier transform in the $X$-, and $Y$-coordinate, then \eqref{eq:KG-asy-dual0} becomes  
\begin{align}\label{eq:KG-asy-xieta}
    &\frac{\xi^2-\eta^2}{\sqrt{\big|1-\frac{\xi^2+\eta^2}{2}\big|}}\left(\partial_\xi^2\chi-\partial_\eta^2\chi\right)\\
    &+\left(\frac{1}{\xi+\eta}+\frac{1}{\xi-\eta}\right)\partial_\xi\chi+\left(\frac{1}{\xi+\eta}-\frac{1}{\xi-\eta}\right)\partial_\eta\chi\nonumber\\
    &+\left(\frac{k^2_X}{(\xi^2-\eta^2)|1-\frac{\xi^2+\eta^2}{2}|}+\left|1-\frac{\xi^2+\eta^2}{2}\right|k^2_Y\right)\chi=0. \nonumber
\end{align}
Although even in this chart, the differential operator will not be separable either, this system is easier tractable because the singularity occurs for all $X$ and $Y$ whenever $\xi^2+\eta^2=2$ and the Klein-Gordon operator \eqref{eq:KG-asy-xieta} retains a symmetry $\xi\leftrightarrow\eta$.

Without loss of generality, we can choose the spatial coordinate $\eta=\eta_o$ and constant, and study only the limit in $\xi$ close to the singularity. Different choices of $\eta_o$ only affect the value for $\xi$ at which the singularity occurs. 
Additionally, for a $\xi$ close enough to the singularity, the spatial motion becomes de facto irrelevant. Hence, the choice $\eta_o=0$ simplifies calculations without spoiling the general argument.

Thus, demanding $\eta_o=0$, we can modify our ansatz for $\chi_{\boldsymbol{k}}(\xi,\eta)\approx\chi_{\boldsymbol{k}}(\xi,0)$ which solves the asymptotic Klein-Gordon equation 
\begin{equation}
    \frac{\xi^2}{\big|1-\frac{\xi^2}{2}\big|^\frac{1}{2}}\frac{{\rm d}^2\chi}{{\rm d}\xi^2}+\frac{2}{\xi}\frac{{\rm d}\chi}{{\rm d}\xi^2}+\left(\!\frac{k^2_X\xi^{-2}}{|1-\frac{\xi^2}{2}|}+\left|1-\frac{\xi^2}{2}\right|k^2_Y\!\right)\chi=0.
\end{equation}
To construct the limit towards the singularity, we further introduce $\upsilon^2:=|1-\frac{\xi^2}{2}|$. The resulting Jacobi determinant yields d$\xi=\frac{2\upsilon}{\sqrt{2-2\upsilon}}$d$\upsilon$ which then leads to the equation
\begin{align}
     \frac{(2-2\upsilon^2)^{\frac32}}{2\upsilon^2}\frac{{\rm d}}{{\rm d}\upsilon}&\left(\frac{\sqrt{2-2\upsilon^2}}{2\upsilon}\frac{{\rm d}\chi}{{\rm d}\upsilon}\right)\\
     &+\frac{1}{\upsilon}\frac{{\rm d}\chi}{{\rm d}\upsilon}+\left(\frac{k^2_X}{\upsilon^2(2-2\upsilon^2)}+\upsilon^2k^2_Y\right)\chi=0\nonumber
\end{align}
where the singularity is located at $\upsilon\to0$. Hence, when performing the approximation $\upsilon\ll1$, we can simplify the above equation tremendously. However, to appreciate the leading singularities, we introduce $\epsilon>0$ as a smallness parameter $\upsilon\to\epsilon\upsilon_o$ with $\upsilon_o$ a constant that carries the dimension. As such, $\partial_\upsilon\to\frac1\epsilon\partial_{\upsilon_o}$, hence, the simplified equation of motion reads after some arrangements
\begin{equation}
    \frac{1}{\epsilon^2}\partial_{\upsilon_o}^2\chi+\left(\frac{1}{\epsilon^2\upsilon_o}+\epsilon\upsilon_o^2\right)\partial_{\upsilon_o}\chi+\left(\epsilon\frac{k^2_X\upsilon_o}{2}+\epsilon^5\upsilon_0^5k_Y^2\right)\chi=0.
\end{equation}
Clearly, the dominant terms are the first two terms which are the only surviving ones for $\epsilon\to0$. The $k_Y^2$-term can be safely neglected here. Up to $\mathcal{O}(\epsilon^2)$, the solution is given by a linear combination of the hypergeometric function and the Meijer G-function
\begin{align}\label{eq:sol_higherorder}
\chi_{k_X}(\xi,0)=c_1 \,& _1F_1\left(\frac{k_X}{3};1;-\frac{|1-\frac{\xi^2}{2}|^{\frac32}}{3}\right)\\
&+c_2 \,\mbox{G}_{1,2}^{2,0}\left(
\begin{array}{c}
 1-\frac{k_X}{3} \\
 0,0 \\
\end{array}\Bigg|\;\frac{|1-\frac{\xi^2}{2}|^{\frac32}}{3}
\right)\nonumber
\end{align}
Considering only terms that are singular in the $\epsilon\to0$ limit, we find a simple $k_X$-independent logarithmic solution 
\begin{equation}
\chi(\xi)=c_1+\frac{c_2}{2}\ln\left|1-\frac{\xi^2}{2}\right|
\end{equation}
This result is of particular interest since, the logarithmic solution is also obtained for the modes in an interior Schwarzschild space-time and a Kasner space-time close to the singularity \cite{hof15,hof19}. 

To perform the quantum probing, we need to explicitly construct the bi-local kernel function \eqref{eq:kernelallg}. The mode-sum decomposition with Huygens's principle and Abel's identity allows to express the kernel in terms of the mode function $\chi_{\boldsymbol{k}}(\xi,0)$ as follows \cite{long1998schrodinger,hof15}
\begin{equation}\label{eq:kernel_equation}
    K_{\boldsymbol{k}}(\xi)=-\frac{i}{\sqrt{-{\rm det}(g)}}\partial_\xi\ln\left(\frac{\chi_{\boldsymbol{k}}(\xi,0)}{\chi_{\boldsymbol{k}}(\xi_o,0)}\right),
\end{equation}
where $\xi_o$ is an arbitrary initial value\footnote{This kernel is related with the Wightman function as being twice its inverse \cite{long1998schrodinger}.}. In our approximation of a fixed $\eta_o\ll\xi$, we find det$(g)\sim-|1-\frac{\xi^2}{2}|^{5}/\xi^4$. Then, taking the solution \eqref{eq:sol_higherorder}, we can derive a near singularity approximation of \eqref{eq:kernel_equation}
\begin{equation}\label{eq:simpelKP-kernel}
  K(\xi)\sim\frac{ i6\sqrt2}{\left|1-\frac{\xi^2}{2}\right|^{7/2} \left(3 \ln \left|1-\frac{\xi^2}{2}\right|-i4\pi\right)}.
\end{equation}
We observe that, in contrast to the folding singularity of a single plane wave \cite{sachs21}, the asymptotic kernel becomes independent of any momentum and any constant $c_1$ or $c_2$. This feature is reminiscent of the solution, we find for Schwarzschild space-time \cite{hof15} and Kasner universes \cite{hof19}\footnote{This apparently universal feature of $k$-independence while approaching the singularity is reminiscent of the BKL-conjecture \cite{BKL1,BKL2} because it suggests that the spatial contribution reduces only to a contact term while the temporal one diverges \cite{hof19}.}. 

In spite of the similarities to the Schwarzschild singularity, the approach towards the singularity and the strength are not exactly identical, which may have consequences for the wave functional. The first step for the quantum probing will be performed through evaluating the normalization \eqref{eq:FSSNNorm} close to the singularity given by the integral of the imaginary part,
\begin{align}
    \ln\left(\frac{|N(\xi)|^2}{|N_0|^2}\right)\sim\mathcal{V}_{\rm PS}&\int_\xi\frac{\sqrt{2}\,\mbox{d}\xi}{|1-\frac{\xi^2}{2}|\ln|1-\frac{\xi^2}{2}|}\\
    &\hspace{2em}\sim\mathcal{V}_{\rm PS}\ln\left(\ln\left|1-\frac{\xi^2}{2}\right|\right)\nonumber.
\end{align}
Here $\mathcal{V}_{\rm PS}=\Xi_{\boldsymbol{k}}\mbox{vol}(\Sigma_\xi)$ denotes the regularized phase space volume, where $\Xi_k$ reflects the regularized momentum contribution and the integral over the spatial hypersurface is denoted by vol$(\Sigma_\xi)$. Note, the value of $|N_0|^2$ is chosen such that the constant of integration cancels. When exponentiated, the normalization function reads
\begin{equation}
\frac{|N(\xi)|^2}{|N_0|^2}\sim\left|1-\frac{\xi^2}{2}\right|^{\mathcal{V}_{\rm PS}}.
\end{equation}
The value of $N(\xi)$ is basically determined by the argument $1-\frac{\xi^2}{2}$ because the phase space volume $\mathcal{V}_{\rm PS}$ is a semi-positively definite object. Its value is defined through the hypersurface volume vol$(\Sigma_\xi)$ and the momentum space volume $\Xi_{\boldsymbol{k}}$. Let us first analyze vol$(\Sigma_\xi)$, to this aim, we consider the behavior of the induced metric essentially provided through a pull back of $\eqref{eq:metrik-xi-eta}$. As one can read off from the metric, the $Y$ component stays finite while the remaining spatial components degenerate to zero. Hence, vol$(\Sigma_\xi)\to0$ for $\xi$ approaching the singularity. In the limit when $\xi$ approaches the singularity 
\begin{equation}
\lim_{\xi\to\sqrt2}|N(\xi)|^2=\lim_{\xi\to\sqrt2}\left|1-\frac{\xi^2}{2}\right|^{\mathcal{V}_{\rm PS}}=1,
\end{equation}
since $\lim_{\xi\to\sqrt2}\mbox{vol}(\Sigma_\xi)=0$. For the exponential term, we evaluate the functional integration over all $\phi\in \mathcal{F}\Sigma_\xi$. This quantity is proportional to the real part of the kernel
\begin{align}\label{eq:funkInt}
\bigg\|\exp(-[\phi]\mbox{Re}(\mathcal{K}(\xi))[\phi])\bigg\|&\!=\!\sqrt{\frac{C}{{\rm Det}[-{\rm det}(g){\rm Re}(K_{\boldsymbol{k}}(\xi))]}}\\
&=\bar C\left|1-\frac{\xi^2}{2}\right|^{-\frac{3A}{2}}\!\!\ln^{A}\left|1-\frac{\xi^2}{2}\right|\nonumber.
\end{align}
We have defined the constants $C$ and $\bar C$ to absorb all numerical factors. Here, the quantity $A$ denotes the number of momentum eigenstates within the regularized momentum space. This quantity can be regularized via a Riemann zeta function: effectively $A$ stems from the $k$ integral in the kernel, therefore, to regularize $A$, we need to regularize $\int$d$k$ which, after discretization, turns our task into regularizing $\sum_{n=1}^\infty 1$. By analytic continuation, $A\to\zeta(0)$ which is $\zeta(0)=-\frac12$. Since, the first term in \eqref{eq:funkInt} dominates for $\xi\to\sqrt2$, we find that norm 
\begin{equation}
\lim_{\xi\to\sqrt2}\|\Psi_\xi[\phi]\|\propto\lim_{\xi\to\sqrt2}\left|1-\frac{\xi^2}{2}\right|^{\frac{3}{4}+\mathcal{V}_{\rm PS}}=0
\end{equation}
In this sense, the quantum probes recover the result in Schwarzschild space-time and the probability for finding quanta near the curvature singularity of the Khan-Penrose space-time approaches zero.

\section{Weakly singular colliding wave solutions}

The Khan-Penrose space-time is one representative of colliding plane-wave space-times; there exists actually a plethora of colliding plane-wave space-times, e.g. Chandrasekhar-Xanthopoulos \cite{chandrasekhar1985colliding}, Bell-Székeres \cite{szekeres1970colliding}, or the Ferrari-Ibáñez space-time \cite{ferrari1987new,yurtsever1988structure}. The latter is of particular interest, since it provides a rich family of different geometries. We will now consider an unusual case, namely, the degenerate Ferrari-Ibáñez solution which features a weak singularity in patch IV. The Penrose diagram can be found in fig. \ref{fig:KPRZl}.

\subsection{Ferrari-Ibáñez space-time: geometry}

The degenerate Ferrari-Ibáñez space-time is found by multiplying $F^2$ with $1/(1-(U^2+V^2))$ and $G^2$ with $1-(U^2+V^2)$ in \eqref{eq:khanpenrose}. This modification changes its nature completely. Considering the Weyl scalars 
\begin{eqnarray}
    \Psi_0&=&\frac{\delta(V)}{(1+U)^2w(U)}-\frac{3}{(1+U\,w(V)+V\,w(U))^3}\nonumber\\
    \Psi_2&=&\frac{1}{(1+U\,w(V)+V\,w(U))^3}\\
    \Psi_4&=&\frac{\delta(U)}{(1+V)^2w(V)}-\frac{3}{(1+U\,w(V)+V\,w(U))^3}\nonumber,
\end{eqnarray}
and comparing them with the Khan-Penrose solution, we immediately see that they admit no longer a singularity at $U^2+V^2=1$ although this divergence is still present in the metric. In fact, all Weyl scalars remain finite over the whole manifold (except for the path where the impulsive waves travel). This suggests that the former curvature singularity has been replaced by a coordinate singularity. In fact, the region $U^2+V^2=1$ describes now a Cauchy horizon\footnote{For a Cauchy surface $\mathscr{S}$, a future Cauchy horizon is defined via $\overline{D^+(\mathscr{S})}-I^+(\mathscr{S})$, that is the future boundary of the future domain of dependence $D^+(\mathscr{S})$ \cite{hawk73}. We defined the closed set $\overline{D^+(\mathscr{S})}:=D^+(\mathscr{S})\cup\partial D^+(\mathscr{S})$, where $\partial D^+(\mathscr{S})$ denoted the boundary.}. We will see, if and in which sense our previous result persists. 

\begin{figure}
    \centering
    \includegraphics[width=\linewidth]{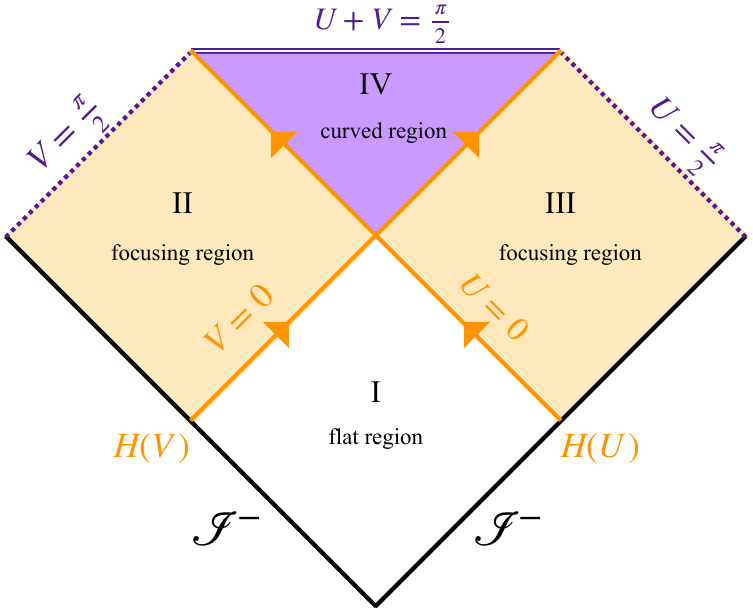}
    \caption{Penrose diagram of the Ferrari-Ibáñez space-time showing two colliding waves at $U=0$ and $V=0$ that cut the space-time into four portions. In the orange shaded, focusing regions, null-geodesics that have crossed the wave are focused at the focal plane $V=\frac\pi2$ in II, or $U=\frac\pi2$ in III. The purple shaded region is curved and admits a Cauchy horizon at $U+V=\frac\pi2$. 
    \label{fig:KPRZl}}
\end{figure}

For the sake of calculations, we will switch to a slightly different parametrization that expresses the metric coefficients through trigonometric functions for which the metric becomes more compact
\begin{eqnarray}
    e^{-M(U,V)}&=&2(1+\sin(p(U)+q(V))),\nonumber\\
    F^2(U,V)&=&\frac{1-\sin(p(U)+q(V))}{1+\sin(p(U)+q(V))},\\
    G^2(U,V)&=&(1+\sin(p(U)+q(V)))^2\cos^2(p(U)-q(V))\nonumber
\end{eqnarray}
where $p(U)$ and $q(V)$ are defined as before. The geometry distinguishes again four patches (cf. Fig. \ref{fig:KPRZl} for the Penrose diagram), while I, II, and III resemble essentially those in the Khan-Penrose space-time. Region IV can still be understood as a curved region; however, now its future is now limited by a Cauchy horizon instead of a singularity.

The above representation of the metric is best-suited to analyze the curved space-time region because IV remains globally hyperbolic. We adapt our description to this via the coordinate transformation \eqref{eq:1+3splitcoords} such that the curved patch admits the metric
\begin{align}\label{eq:metriktrig}
&g=-(1+\sin(\xi))[\mbox{d}\xi\otimes\mbox{d}\xi-\mbox{d}\eta\otimes\mbox{d}\eta]\\
&+\frac{1-\sin(\xi)}{1+\sin(\xi)}\mbox{d}X\otimes\mbox{d}X+(1+\sin(\xi))^2\cos^2(\eta)\mbox{d}Y\otimes\mbox{d}Y.\nonumber
\end{align}
As before, the coordinates induce a natural $(1+3)$-decomposition where now $0\le\xi<\frac\pi 2$ and $-\frac\pi 2\le\eta\le\frac\pi 2$, while $X$ and $Y$ remain the same. In this parametrization, the geodesic boundary is located at $\xi=\frac\pi2$ if we set $\eta=\eta_o=0$ as before. However, this boundary is not realized as a curvature singularity but a Cauchy horizon. This is obvious because none of the Weyl scalars diverge with the consequence that the Kretschmann scalar \eqref{eq:Kretschmann} remains finite as well.

\subsection{Ferrari-Ibáñez space-time: quantum probing}

The quantum probing of the degenerate Ferrari-Ibáñez solution involves very similar steps as the previous section. Therefore, we will report only the important steps and refer to the relevant equations if necessary. To solve the Klein-Gordon equation in the curved patch, we work in the coordinate chart \eqref{eq:metriktrig}, for which the d'Alembert operator becomes separable such that
\begin{equation}\label{eq:modensumme}
    \phi(\xi,\eta,X,Y)=\int\frac{{\rm d}^2\boldsymbol k}{4\pi^2}\sum_\alpha \psi_\alpha(\xi;\boldsymbol{k})\varrho_\alpha(\eta;\boldsymbol{k})e^{i\boldsymbol{k}\boldsymbol{x}},
\end{equation}
where we defined the separation constant $\alpha$, plus the two-dimensional vectors $\boldsymbol{k}=(k_X,k_Y)^T$ and $\boldsymbol{x}=(X,Y)^T$. This feature simplifies calculations tremendously compared to the non-separable Khan-Penrose d'Alembert operator. The two ordinary differential equations for $\psi_\alpha(\xi;\boldsymbol{k})$ and $\varrho_\alpha(\eta;\boldsymbol{k})$ read \cite{yurtsever1989quantum}
\begin{eqnarray}
    \frac{{\rm d}^2\psi_\alpha}{{\rm d}\xi^2}(\xi;\boldsymbol{k})&-&\tan(\xi)\frac{{\rm d}\psi_\alpha}{{\rm d}\xi}(\xi;\boldsymbol{k})\label{eq:psialpha}\\
    &&+\left(\alpha+\frac{k_X^2(1+\sin(\xi))^4}{\cos^2(\xi)}\right)\psi_\alpha(\xi;\boldsymbol{k})=0,\nonumber\\
    \frac{{\rm d}^2\varrho_\alpha}{{\rm d}\eta^2}(\eta;\boldsymbol{k})&-&\tan(\eta) \frac{{\rm d}\varrho_\alpha}{{\rm d}\eta}(\eta;\boldsymbol{k})\label{eq:phialpha}\\
    &&+\left(\alpha-\frac{k_Y^2}{\cos^2(\eta)}\right)\varrho_\alpha(\eta;\boldsymbol{k})=0\nonumber.
\end{eqnarray}
Since, we are particularly interested in the vicinity of the Cauchy horizon, i.e. around $\xi\approx\frac\pi2$, we can expand the trigonometric functions in \eqref{eq:psialpha} to $\mathcal{O}(\xi-\frac\pi2)$ such that we end up with the asymptotically expanded differential equation
\begin{align}\label{eq:psiAEx}
        \frac{{\rm d}^2\psi_\alpha}{{\rm d}\xi^2}(\xi;\boldsymbol{k})&+\frac{1}{\xi-\frac\pi2} \frac{{\rm d}\psi_\alpha}{{\rm d}\xi}(\xi;\boldsymbol{k})\\
        &+\left(\frac{16k_X^2}{(\xi-\frac\pi2)^2}+\alpha-\frac{32k_X^2}{3}\right)\psi_\alpha(\xi;\boldsymbol{k})=0.\nonumber
\end{align}
Clearly, the above equation is a Bessel differential equation and, therefore, we find as solution the Bessel function of the first and second kind
\begin{align}\label{eq:psiBessel}
    \psi_\alpha(\xi;\boldsymbol{k})=&c_1\mbox{J}_{i4k_X}\left(-\frac{i}{3}\sqrt{96k^2_X-9\alpha}\left(\xi-\frac\pi2\right)\right)\\
    &+c_2\mbox{Y}_{i4k_X}\left(-\frac{i}{3}\sqrt{96k^2_X-9\alpha}\left(\xi-\frac\pi2\right)\right).\nonumber
\end{align}
To get a better understanding of their behavior at the singularity, we consider a small argument expansion, or alternatively perform an asymptotic expansion and only evaluate \eqref{eq:psiAEx} for singular coefficients; both cases yield
\begin{equation}\label{eq:psiexpfunk}
    \psi_\alpha(\xi;\boldsymbol{k})\sim c_1\,e^{i4k_X\ln(\frac\pi2-\xi)}+c_2\,e^{-i4k_X\ln(\frac\pi2-\xi)}.
\end{equation}
This solution is familiar from the Schwarzschild horizon where, due to the coordinate singularity of the Schwarzschild coordinates, the wave gets infinitely blue-shifted. Since, the Ferrari-Ibáñez singularity manifests itself as a Cauchy horizon, we observe a similar blue-shifting in the exponent. On the other hand, this suggests that we can continue this solution across the Cauchy horizon by analytically continuing the logarithmic term $i4k_X\ln(\frac\pi2-\xi)\to i4k_X\ln|\xi-\frac\pi2|\pm4\pi k_X$ which is analogous to Unruh's prescription \cite{unruh1976notes}. 
To close this analysis of the Klein-Gordon equation, let us present the solution to equation \eqref{eq:phialpha} for $\varrho_\alpha(\eta;\boldsymbol{k})$, that is,
\begin{align}
   \varrho_\alpha&(\eta;\boldsymbol{k}) = \!\frac{z_1\, _2F_1\left(\frac{B^1_-}{4},\frac{B^1_+}{4} \left(1-2 k_Y+\beta\right);1-l;\cos ^2(\eta)\right)}{i^{k_Y} \cos ^{k_Y}(\eta)}\\
   &+i^{k_Y} z_2
    \cos ^{k_Y}(\eta) \, _2F_1\left(\frac{B^2_-}{4},\frac{B^2_+}{4}
;l+1;\cos ^2(\eta)\right)\nonumber
\end{align}
where we introduced $\beta=\sqrt{4 \alpha+1}$, $B^1_\pm=1-2 k_Y\pm\beta$, and $B^2_\pm=1+2 k_Y\pm\beta$ for the ease of notation. It has been shown that, when properly normalized, these modes form a complete orthonormal basis \cite{yurtsever1989quantum}
\begin{equation}
    \int_{-\frac\pi2}^\frac{\pi}{2}\mbox{d}\eta\cos(\eta)\varrho_\alpha(\eta;k_Y)\varrho^*_{\alpha'}(\eta;k_Y')=\delta_{\alpha\alpha'}\delta(k_Y-k_Y').
\end{equation}
Since there is no particularly interesting limit associated to the $\eta$-coordinate, we close the discussion here. 

In the following, we construct the bi-local kernel function $K_{\boldsymbol{k}}(\alpha;\xi)$. As before, we express the kernel via \eqref{eq:kernel_equation} in terms of the mode function $\psi_\alpha(\xi;\boldsymbol{k})$. Since we will work in the vicinity of the singularity, we use the asymptotic mode solution in \eqref{eq:psiAEx} and find for the bi-local kernel function 
\begin{equation}\label{eq:kxi}
      K_{\boldsymbol{k}}(\alpha;\xi)=\frac{k_X}{(\xi-\frac\pi2)^2\cos(\eta)}\left(\frac{2c_2}{c_2+c_1(\xi-\frac\pi2)^{i8k_X}}-1\right).
\end{equation}
As we see, $K_{\boldsymbol{k}}(\alpha;\xi)$ diverges quadratically at the Cauchy horizon and remains $k_X$-dependent. Note, to get \eqref{eq:kxi}, we expanded the metric determinant close to the value $\xi=\frac\pi2$ which yields 
\begin{equation}\label{eq:deter}
    \sqrt{-\mbox{det}(g)}=(1+\sin(\xi))^2\!\cos(\xi)\cos(\eta)\sim4\!\left(\frac\pi2-\xi\right)\!\cos(\eta).
\end{equation}
Let us now study how this inflicts the dynamics of the field configurations in the wave-functional. In particular, we ask whether the field configurations can migrate onto or even across the $\xi=\frac\pi2$ hypersurface. 

We start with the derivation of the normalization function $|N(\xi)|^2$. As follows from \eqref{eq:FSSNNorm}, this quantity involves an integral of the imaginary part of the kernel. Evaluated in the vicinity of the Cauchy horizon, the integral assumes the following form 
\begin{align}\label{eq:FINormal}
    \int\mbox{d}^4x&\int_k\frac{{\rm d}\boldsymbol k}{(2\pi)^2}\sqrt{-\mbox{det}(g)}\,\mbox{Im}(K_{\boldsymbol{k}}(\alpha;\xi))\\
    &=\frac{\bar C_2\mathcal{V}_{\rm PS}}{4\pi^2}\int_\xi\frac{\mbox{d}\xi}{\xi-\frac\pi2}=\frac{\bar C_2\mathcal{V}_{\rm PS}}{4\pi^2}\ln\Big|\xi-\frac\pi2\Big|.\nonumber
\end{align}
In combination with the measure, \eqref{eq:FINormal} amounts to a logarithmic dependence, thus, rendering $|N_\xi|^2\propto|\xi-\frac\pi2|^{\bar C_3\mathcal{V}_{\rm PS}}$. Its behavior is almost fully determined by the quantities in the phase-space volume, that is, the momentum space parameter $\Xi_k$ and the volume of the hypersurface vol$(\Sigma_\xi)$. For simplicity, we choose $\eta=0$ without loss of generality. 

Let us first discuss the volume of the hypersurface. For this quantity, we evaluate its limit towards $\xi=\frac\pi2$ by considering a unit volume on $\Sigma_\xi$. For $\xi\to\frac\pi2$, the $\eta$- as well as the $Y$-direction assume a finite value, the $X$-direction shrinks to zero, thus the unit volume is not conserved and becomes eventually zero, i.e. $\lim_{\xi\to\frac\pi2}\mbox{vol}(\Sigma_\xi)=0$ which is similar for plane-wave space-times and, thus, for the portions II and III. Since vol$(\Sigma_\xi)\to0$, we may conclude $|N(\xi)|^2\to1$ at the Cauchy horizon. 

To fully sustain this verdict, we need to analyze the momentum space quantity $\Xi_k$. Since there is a choice for the constants of integration, they can in principle admit a $\boldsymbol{k}$-dependency. Assuming this were not the case, $\Xi_k$ contains the integral over $k_X$. Since $k_X$ is an odd function integrated over a symmetric interval, we can conclude that $\Xi_k$ vanishes. This is consistent with the popular choice of the vacuum in this space-time which sets $c_1\equiv0$ and $c_2\in\mathbb{R}$ in \eqref{eq:psiexpfunk} \cite{yurtsever1989quantum,ash75}. On the other hand, the momentum space volume remains independent of any coordinate value, thus its value is independent of the hypersurface and may therefore be ignored. As a consequence, the kernel function admits no imaginary part, thus the integral yields zero and $N(\xi)\to1$. Therefore, the norm for any field configuration to reach the singularity at $\xi=\frac\pi2$ becomes one.

Moving to the second contribution, we perform the functional over the field configurations $\phi\in \mathcal{F}\Sigma_\xi$ which yields for the asymptotic kernel 
\begin{align}
   \bigg\|&\exp\left(-[\phi]\mbox{Re}(\mathcal{K}(\xi))[\phi]\right)\bigg\|\\
   &\hspace{1em}=\left|\frac{C}{\mbox{Det}[-\mbox{det}(g)\mbox{Re}(K(\xi))]}\right|^\frac{1}{2}\propto\mbox{Det}^{-\frac12}\left(\bar C_1k_X\right),\nonumber
\end{align}
with $C$ and all (upcoming) $\bar C$ collecting the appearing constants. As we can immediately observe, the $\xi$-dependence in the determinant \eqref{eq:deter} cancels the one in \eqref{eq:kxi} exactly, leaving an effectively constant functional determinant. The functional determinant is evaluated via the $\zeta$-function method which for any quantity $\mathcal{Q}$ with eigenvalues $\lambda_n$ \cite{SachsWipf,zerb02}
\begin{equation}\label{eq:zeta}
    \mbox{Det}(\mathcal{Q})=\prod_n\lambda_n=\left.\exp\left(-\frac{{\rm d}}{{\rm d}s}\zeta(s;\mathcal{Q})\right)\right|_{s=0},
\end{equation}
with $\zeta(s):=\sum_n\frac{1}{\lambda_n^s}$. Notice, Re$(s)>1$ for the sum to converge. Combining the two formulae shows that the functional determinant can be understood as a product of the eigenvalues $\bar C_1 k_X$. From the explicit definition of the $\zeta$-function in  \eqref{eq:zeta} follows that
\begin{align}
    \zeta(s)&=\frac{1}{2\pi}\int_{-\infty}^{\infty}\frac{{\rm d}k_X}{(\bar C_1k_X)^s}\\
    &=\lim_{k_X\to\infty}-\frac{k_X((\bar C_1k_X)^s+(-\bar C_1k_X)^s)}{(1-s)(-\bar C_1k_X)^{2s}}=0.\nonumber
\end{align}
The last equation only holds under the condition $\mbox{Re}(s)>1$ which is consistent with the condition for the $\zeta$-function itself. Since the $\zeta$-function is zero, so will be the derivative. With \eqref{eq:zeta} follows then immediately that the functional determinant becomes one in the vicinity of the Cauchy horizon. Hence, 
\begin{equation}
    \lim_{\xi\to\frac\pi2}\|\Psi_\xi[\phi]\|=1,
\end{equation}
for the degenerate Ferrari-Ibáñez solution. As a consequence, the quantum field will reach the Cauchy horizon and may be able to pass through (when backreactions are neglected). 

\section{Classification and consequences}

Collecting the results from different patches of the colliding wave space-time and previous analyses of the Schwarzschild \cite{hof15} and Kasner space-times \cite{hof19}, we observe an interesting pattern when quantum probes approach a singularity. All three space-times, patch IV in Khan-Penrose space-time, Schwarzschild space-time, and Kasner space-time admit a complete evolution such that no field configuration can reach the geodesic border. In contrast, the folding singularities in patch II and III as well as the spacelike singularity in patch IV of the Ferrari-Ibáñez space-time can be reached.

\subsection{The $\Psi_2$-conjecture}

Although all the complete space-times differ in symmetry, one important similarity marks the divergence of the Coulomb part of the Weyl tensor $\Psi_2$. While for the Khan-Penrose patch, $\Psi_0$, $\Psi_2$ as well as $\Psi_4$ diverge at the singularity, $\Psi_2$ is the only non-vanishing contribution in the Schwarzschild interior. Since, there, $\Psi_2$ diverges as well, this suggests a pattern. The analyses of the folding singularities in the patches II and III agree with the result for the single plane wave in \cite{sachs21}, i.e. the field configurations will be able to reach the focal plane. For both patches $\Psi_2\equiv0$ while only one of the transversal components, either $\Psi_0$ or $\Psi_4$ diverge while the respective other vanishes. 

This shows that the transversal components $\Psi_0$ and $\Psi_4$ albeit divergent, do not inflict a depletion of wave-functional's norm but a divergent $\Psi_2$ will. This is substantiated by the result from the Ferrari-Ibáñez space-time where none of the Weyl scalars diverges in patch IV and the wave-functional is able to propagate the field configurations onto the geodesic boundary. At this point, we would like to formulate our conjecture that determines the evolution of quantum field in singular space-times (with non-vanishing Weyl curvature).

\begin{remark}[Quantum Weyl Conjecture]
    The evolution of quantum probes is shielded from the curvature singularity, if the singularity manifests as a divergence of the Penrose-Newman Weyl scalar $\Psi_2$.
\end{remark} 

Since the $\Psi_2$ Penrose-Newman scalar is commonly referred to as the Coulomb part of the curvature, this interpretation enforces the idea of a gravitational hydrogen atom for a spacelike singularity. In both cases, the Coulomb part of the potential inflicts a probabilistic obstacle shielding the quantum degree of freedom from the singularity. Note, due to considering only matter free space-times, there may be the caveat that also Ricci Penrose-Newman scalars\footnote{Cosmological Friedmann-Lemaître-Robertson-Walker space-times have not yet been analyzed within the functional Schrödinger approach but judging from the Heisenberg picture, the singularity can be passed \cite{ashtekar2021probing,Ashtekar:2022oyq} even for arbitrary sectional curvature \cite{Ashtekar:2022cih}.} contribute, but at least for the Khan-Penrose space-time, the important contributions
\begin{eqnarray}
    \Lambda&=&\frac{R}{24}=\frac{e^M}{24}(\partial_U\Omega\partial_V\Omega-2\partial_U\partial_VM-2\partial_U\partial_V\Omega)\nonumber\\
    \Phi_{11}&=&\frac14 (Ric(l,n)+Ric(m,\bar m))\\
    &&=\frac14\partial_U\partial_VM+\frac{3}{16}\partial_U\partial_V\Omega-\frac{3}{32}\partial_U\Omega\partial_V\Omega\nonumber\,,
\end{eqnarray}
with both, $\Lambda\supset\Psi_2$ and $\Phi_{11}\supset\Psi_2$; hence, they contribute similarly. The other Penrose-Newman scalars related to the Ricci tensor contain only one type of derivative, hence, they are reminiscent of $\Psi_0$ or $\Psi_4$. 

One may wonder, why $\Psi_2$ is more significant than $\Psi_0$ and $\Psi_4$. The answer lies in the invariance of these objects. While $\Psi_2$ is Lorentz invariant, $\Psi_0$ and $\Psi_4$ have boost weight $-2$ and $+2$. Hence, a divergence in one of the two can be suppressed by a sufficiently strong boost. A divergence of the Coulomb part given by $\Psi_2$ cannot be mitigated through a Lorentz transformation and to our knowledge, we could not find a space-time that features a divergent $\Psi_0$ or $\Psi_4$ while keeping a finite $\Psi_2$.

\subsection{Backreaction}\label{sec:5}
Since colliding plane waves provide a (sort of) idealization of high energy scattering on a single plane wave (which in turn can be thought of as an idealization of a trans-Planckian 2 $\to$ 2 scattering), it would be interesting to analyze whether taking backreaction into account, one can interpolate between the folding singularity of a single plane wave and the space-like singularity of the  Khan-Penrose space-time.  

A convenient approach is to exploit the semiclassical Einstein equation $G=8\pi \langle T\rangle_{\rm ren}$ that relates the Einstein tensor with the stress-energy tensor. Already this system imposes severe technical difficulties, hence, we rely on a more pragmatic approach using our conjecture. The following approach should be understood as a classical (or quasi-classical) estimate of the backreactions. A full determination would require specifying the quantum state and performing a full stress-energy tensor renormalization procedure. However, we show how the classical stress-energy tensor provides a guideline given a natural choice for the vacuum state in a pp-wave spacetime. We start with the explicit form for the stress-energy tensor of a free, massless scalar field
\begin{eqnarray}\label{eq:SETallg}
    T=\nabla\phi\otimes\nabla\phi-\frac g2\;g(\nabla\phi,\nabla\phi)
\end{eqnarray}
in a given background. Based on our previous findings, we are interested which contributions of the stress-energy tensor become relevant to understand the need for backreaction when approaching the singularity. It should be understood that the expectation value $\langle T_{ab}\rangle_{\rm ren}$ of the renormalized stress-energy tensor differs in general from \eqref{eq:SETallg} due to normal ordering that involves a choice of a vacuum.

\begin{figure}    
\centering
    \includegraphics[width=\linewidth]{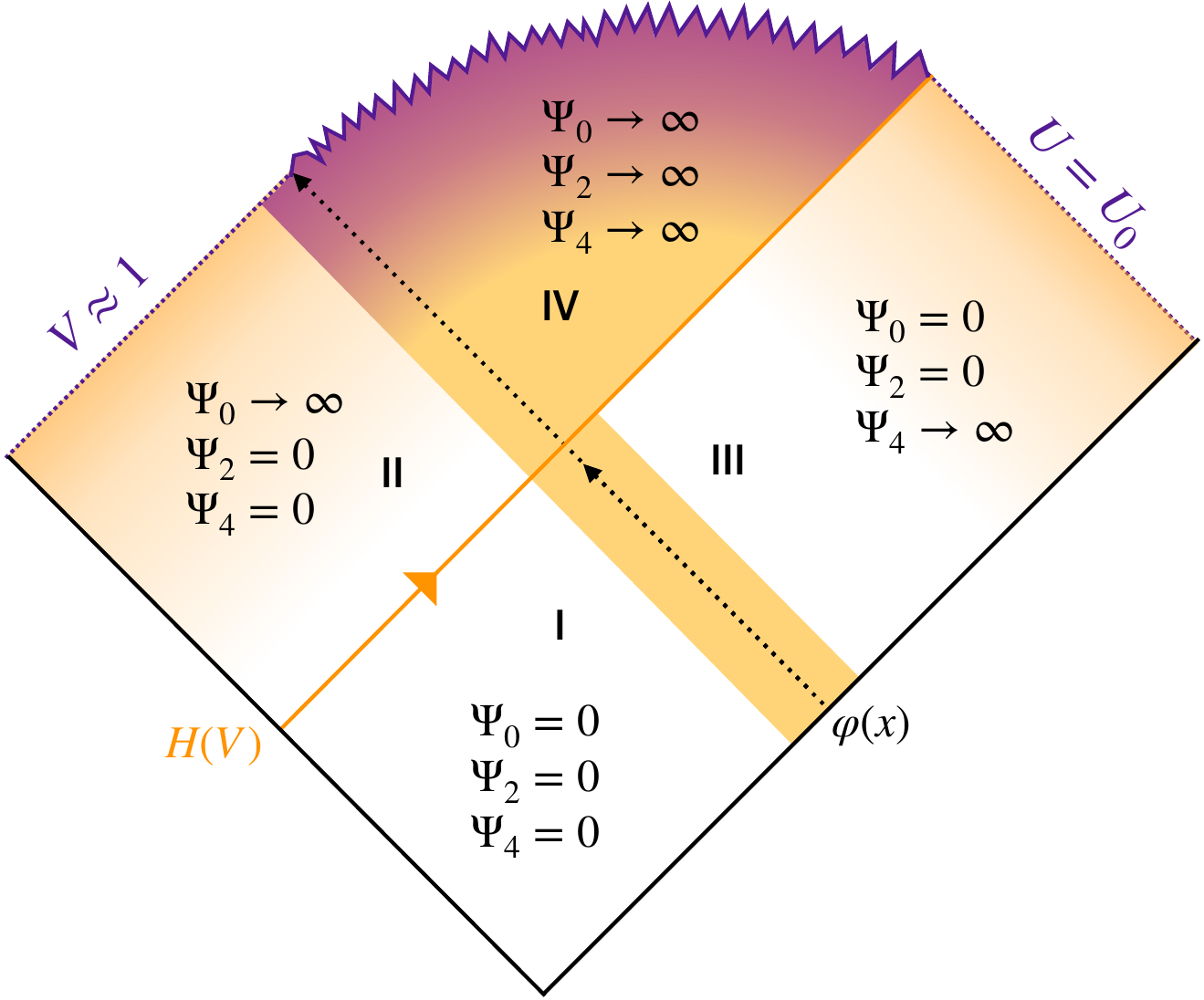}
    \caption{The null-field wave packet $\varphi(x)$ (depicted as the orange bar) is radiated across an impulsive plane-wave $H(V)$ into the region II. After crossing the plane-wave, the field scatters and creates a non-vanishing, actually diverging, $\Psi_2$-Weyl scalar through backreaction. The folding singularity in II transforms into a spacelike singularity and, therefore, into a Khan-Penrose space-time. The focusing that occurs through the wave packet itself may lead to a folding singularity at some $U_0\gg1$. }
  \label{fig:incomingfield}
\end{figure}

In order to probe the stability of the folding singularity, we send a field $\varphi(x)$, representing an $in$-state with respect to the vacuum, that is not per se a test field, from region I through the plane wave $H(V)$ (cf. fig. \ref{fig:incomingfield}). As an initial metric, we use the plane-wave metric that describes patch I and II in fig. \ref{fig:KPRZl}
\begin{equation}
    g_{\rm pw}=-2\mbox{d}U\otimes\mbox{d}V+\gamma_{\rm pw}(V).
\end{equation}
Since the pp-wave propagates along $V=0$, the geometry focuses in the $X$-direction along the $V$-coordinate.

Then, sending a $\varphi(x)$ experiences a space-time of the form of patch $II$ after crossing $H(V)$. We can obtain a physical intuition of the back reaction recalling that $T_{XX}$ represents the (anisotropic) stress component in the $X$-direction. Since this direction collapses due to focusing, $T_{XX}$ diverges, thus triggering a back-reaction.
Alternatively, one could also interpret the scalar profile as an incoming wave itself, with profile $H(U)$. Hence, for the backreacted metric, we make the particular ansatz \cite{hadjiivanov1979free,yurtsever1988colliding}
\begin{equation}\label{eq:stoyanov}
    \tilde g_{\rm BR}=-2e^{-M(U,V)}\mbox{d}U\otimes\mbox{d}V+e^{\Omega(U,V)}\mbox{d}X\otimes\mbox{d}X+\mbox{d}Y\otimes\mbox{d}Y.
\end{equation}
We could also choose a more general form of the space-time, to describe an inhomogeneous profile of the incoming scalar, but the above form serves best for illustrational purposes. Derivation of the curvature tensors yields the following non-vanishing components of the Einstein tensor
\begin{eqnarray}
G_{UU}&=&2\partial_UM\partial_U\Omega-4(\partial_U\Omega)^2+2\partial_U^2\Omega,\\
G_{VV}&=&2\partial_VM\partial_V\Omega-4(\partial_V\Omega)^2+2\partial_V^2\Omega,\\
G_{UV}&=&4\partial_V\Omega\partial_U\Omega-2\partial_V\partial_U\Omega,\\
G_{XX}&=&e^{M-4\Omega}\partial_U\partial_VM,\\
G_{YY}&=&-e^M(8\partial_U\Omega\partial_V\Omega-\partial_U\partial_VM-4\partial_U\partial_V\Omega)\,.
\end{eqnarray}
The backreacted metric features a $U$-dependence taking into account the influence of the field itself, but how consistent is such an assumption? The answer depends not only on local curvature invariants, but also on global properties, like the particular vacuum state. Gibbons \cite{gibbons1975quantized} argued that there exists a natural choice for the vacuum in such space-times, that is an ingoing vacuum state, with respect to a scalar field that propagates along a Killing direction. For such a vacuum, the expectation value of the renormalized stress-energy tensor does not contribute \cite{gibbons1975quantized,zauderer1971modification,friedlander1975wave}. 
Albeit particular, this setup seems natural to understand how an ingoing null field backreacts in a pp-wave space-time.

To understand the contribution from the incoming scalar, let us assume a (weak) $U$-dependence for the field $\varphi(x)$\footnote{This is a generalization of the null-reduction, performed in \cite{sachs21}.}. Then, a plane-wave with $U$-dependence
\begin{equation}
    \varphi(x)=\int_{\vec{k}}\frac{\mbox{d}^3k}{(2\pi)^3}f_{\vec{k}}(V)e^{-ik_UU}e^{i\boldsymbol{k}\boldsymbol{x}},
\end{equation}
will imply a non-zero $T_{UU}$-component $T_{UU}\propto k_U^2$ by \eqref{eq:SETallg}. The Einstein equation then implies that a non-vanishing $T_{UU}$ sources a non-vanishing $G_{UU}$. Since $G_{UU}$ contains only terms proportional to $\partial_U\Omega$ and $\partial_UM$, the backreacted metric must then depend explicitly, but weakly, on $U$. In other words, even the slightest $U$-dependence of $\varphi(x)$ will bring the system closer to a Khan-Penrose-type  scenario. So far we have only shown that $\Omega(U,V)$ carries an explicit $U$-dependence. 

Recalling the form of the solution to the Klein-Gordon equation on a plane-wave background
\cite{garr91,sachs19} 
\begin{equation}\label{eq:fieldansatz}
    \varphi(x)=\frac{2}{\sqrt[4]{{\rm det}(\gamma)}}e^{-ik_UU}e^{i\boldsymbol{kx}}\,,
\end{equation}
where we have defined the spatial metric $\gamma$ to be of the form
\begin{equation}
    \gamma=\begin{pmatrix}
        e^{4\Omega(U,V)}&0\\0&1
    \end{pmatrix}\
\end{equation}
In the above equation, we used the fact, that the metric function $\Omega$ is both $U$ and $V$-dependent. This implies an interesting consequence, because then, the $T_{XX}$- and $T_{YY}$-components are non-vanishing:
\begin{eqnarray}
    T_{XX}&=&e^{M-2\Omega}\partial_U\Omega\partial_V\Omega,\\
T_{YY}&=&e^{M+2\Omega}\partial_U\Omega\partial_V\Omega,
\end{eqnarray}
which in turn requires $G_{XX}$ and $G_{YY}$ to be non-vanishing. Analyzing the form of $G_{XX}$, we see that it scales like $e^{-2\Omega}\Psi_2$. The focusing singularity of the initial metric has been chosen to occur in the $XX$-component, hence, $\Omega\to-\infty$ at the folding singularity $V_0$. Here, $V_0$ is close to $1$ with a small deviation implied by the weak $U$-dependence of $\Omega$. Since, $T_{YY}=e^{4\Omega}T_{XX}$, this contribution is suppressed by an exponential factor of $\Omega$, therefore, we may assume that $G_{YY}$ remains subdominant. 

Knowing that $T_{XX}$ diverges then implies that the $G_{XX}$ component of the Einstein tensor diverges. As a consequence, the metric function $M$ is actually a function of $U$ and $V$. Furthermore, 
\begin{align}\label{eg:breaquation}
    \Psi_2=&e^{2\Omega}G_{XX}=-8\pi e^{2\Omega}T_{XX} \\=&-8\pi e^{M(U,V)-2\Omega(U,V)}\partial_U\Omega(U,V)\partial_V\Omega(U,V)
\end{align}
diverges, since $\partial_V\Omega$ diverges at the folding singularity. 
Hence, a slight $U$-dependence of a scalar field in an impulsive pp-wave background creates a strong singularity with divergent $\Psi_2$, which in turn, by our conjecture, shields the field from populating this singularity, thus, implying again a quantum-field theoretically complete behavior. In this sense, once quantum fields are focused in a null-singularity, they will deform the singularity immediately such that the interplay with gravity protects quantum field theory from reaching the singularity. 

In summary, backreaction can be effectively understood from the point of view of the Weyl quantum conjecture because $\Psi_2$ is the relevant object for estimating the occurrence and strength of backreaction effects.

\section{Discussion}

Plane-wave space-times are of significant interest since every space-time provides a plane-wave limit \cite{blau2004penrose}, when sufficiently zoomed in. In this sense, they are extraordinarily versatile to estimate statements that can be attributed to local processes. A formidable application of this limit is the link between the singular structure of the colliding wave region and the Kasner singularities which in a limiting case mimic the Schwarzschild singularity. 
Geometrically, they admit folding, null singularities which turn into spacelike singularities after the waves collide. Hence, there are two distinct regions, those before the collapse and the one after. Hence, colliding plane-wave space-times offer an ideal system to retrieve information about what determines how quantum probes approach a singularity.

In this article, we focused on the curved space-time region that is created due to the scattered fields after the collision. The two models we considered, were the Khan-Penrose space-time and a slight modification known as the degenerate Ferrari-Ibáñez space-time. Both space-times were picked to illustrate the different behaviors quantum probes may admit. 

For the Khan-Penrose space-time, we found that the curvature singularity is reminiscent of the Schwarzschild singularity and prohibits the quantum fields to reach the singular hypersurface\footnote{A geometrically inspired proposal that uses a changing vacuum can be found in  \cite{alonso2024geometric,alonso2024geometric1}.}. In contrast, the Ferrari-Ibáñez solution yields that any field configuration is able to reach the singular hypersurface. The key to understand this difference was found by analyzing the singular structure which in this space-time is given by a weak singularity as in the regions before the collision but after crossing one wave. Quantum probes exhibit similar behavior in patches II and III due to the absence of significant obstructions to field configurations. The normalizability of the wave-functional remains intact, allowing fields to propagate through the focusing planes. This behavior extends to the curved region of the Ferrari-Ibáñez space-time. Geometrically, the divergence at the Cauchy horizon resembles the one at the Schwarzschild horizon. Knowing how the space-time continues beyond the Cauchy horizon ensures that fields will propagate into this region as well. 

Let us understand the generality of our analysis of a massless scalar field theory by discussing different setups: A mass term will always be subdominant close to the singularity due to a metric determinant as a prefactor which vanishes towards the singularity. This makes sense, since the mass is an infrared quantity while kinetic terms dominate the behavior at the singularity. For spinor fields, we would expect a similar behavior as  for scalar field because the Khan-Penrose space-time resembles Schwarzschild black holes, thus, the differences due to spin and charge will not be present. For higher spin gauge fields, the analysis would feature additional constraints from the gauge condition, e.g. the Gupta-Bleuler condition $\nabla E=0$ applied to $\Psi[a]$ becomes $\nabla \frac{\delta}{\delta a}\Psi[a]=0$ with $a$ the physical degree of freedom \cite{Schneider:2018tyr}. However, it has been shown that spin does not avert singularities \cite{stewart1973can}, thus, we regard the scalar analysis to be sufficient.

Embedding our findings into the landscape of quantum completeness shows that the analogy of the gravitational hydrogen atom might be more than a mere intuitive picture. In fact,  considering the results for the Schwarzschild interior and comparing them with the colliding plane wave space-times, unveils a pattern. In both space-times, the curved region is qualified by a diverging Penrose-Newman scalar $\Psi_2$ that describes the Coulomb part of the Weyl tensor. Taking the analogy serious, quantum completeness seems to be inferred by a diverging $\Psi_2$ rather than any other Weyl Penrose-Newman scalar. Comparison with the plane-wave patches II and III shows that none of the other non-zero, and even diverging, scalars $\Psi_0$ or $\Psi_4$ lead to a depletion of the norm. The Ferrari-Ibáñez solution also substantiates that a divergence in $\Psi_2$ is necessary rather than any finite contribution. 

Through our conjecture, we were able to construct a versatile path of considering backreactions in a specific but natural scenario in which the effect can be described as quasi-classical. 
In fact, we identified the dominant contribution that triggers backreaction. 
Such field configurations mimic a second plane-wave profile that crosses the classical wave, such that the folding, null-singularity "bends" into a spacelike singularity.  From the standpoint of quantum field theory, this feature is beneficial: the singularity is not operationally accessible, and consistency is maintained through backreaction on the geometry, which induces a Coulomb-like Weyl curvature component. We would like to stress that if we had chosen a generic in-vacuum, the vacuum fluctuations would contribute to the backreaction through the semi-classical Einstein equation as well. Such a scenario has been studied e.g. in \cite{yurtsever1989quantum} which required to compute the expectation value of the stress-energy tensor. 

\begin{acknowledgments}
We would like to thank Santiago Londoño Castillo and Simon Kuhn for early  collaboration. MS wants to thank the LMU Munich and the Arnold-Sommerfeld Center for Theoretical Physics for the hospitality during the finalization of the project. Thank you to the anonymous referee for their constructive and detailed comments.
This work is supported by the Excellence Cluster Origins of the DFG under Germany's Excellence Strategy EXC-2094 390783311 as well as EXC 2094/2: ORIGINS 2 and by the Basque Government Grant
\mbox{IT1628-22} as well as by the Grant PID2021-123226NB-I00 (funded by
MCIN/AEI/10.13039/501100011033 and by ``ERDF A way of making Europe'').
\end{acknowledgments}
\bibliography{aapmsamp}

\end{document}